\documentclass{ws-procs975x65}
\begin{document}
\title{Charmful baryonic $B\to {\bf B\bar B'}M_c$ decays}
\author{Yu-Kuo Hsiao$^{*}$}
\address{Institute of Physics, Academia Sinica,\\
Taipei, Taiwan 115, R.O.C.\\
$^*$E-mail: ykhsiao@phys.sinica.edu.tw}
\begin{abstract}
We study the charmful three-body baryonic $B$ decays with $D^{(*)}$ or $J/\Psi$ in the final state. We explain the measured rates of $\bar
B^0\to n\bar p D^{*+}$, $\bar B^0\to p\bar p D^{(*)0}$, and $B^-\to
\Lambda \bar p J/\Psi$. In particular, the branching fraction of $\bar B^0\to\Lambda\bar \Lambda D^0$ predicted to be of order $2.3\times 10^{-6}$ is
in accordance with the Belle measurement,
$(1.05^{+0.57}_{-0.44}\pm 0.14)\times 10^{-5}<2.6\times 10^{-5}$.
\end{abstract}

\section{Introduction}
The experimental measurements present that the charmful and charmless three-body baryonic $B$ decays have the same features.
First, the threshold effect is observed in $\bar B^0\to p\bar p D^{(*)0}$ \cite{ppD(star)_Babar},
where a curve peaks near the threshold area in the dibaryon invariant mass spectrum.
This phenomenon found in all charmless cases can be understood in terms of a simple
short-distance picture \cite{Suzuki}.
Particularly, this picture has been used to realize why charmless three-body decays
\cite{Lambdappi_Belle,LambdaLambdaK_Belle,ppK_Babar,AD_Belle,Lambdapbargamma_Belle,ADLambdapbarpi_Belle,ppKpi_Belle}
have rates larger than charmless two-body decays \cite{BtoBB_Babar,BtoBB_Belle}; that is, $\Gamma(B\to {\bf B \bar B'}
M)>\Gamma(B\to{\bf B\bar B'})$.
One energetic $q\bar q$ pair must be emitted back to back by a hard gluon in order to produce a
baryon and an antibaryon in the two-body decay. This hard gluon is
highly off mass shell and hence the two-body decay amplitude is
suppressed by order of $\alpha_s/q^2$. In the three-body baryonic
$B$ decays, a possible configuration is that the ${\bf B\bar B'}$
pair is emitted collinearly against the meson. The quark and
antiquark pair emitted from a gluon is moving nearly in the same
direction. Since this gluon is close to its mass shell, the
corresponding configuration is not subject to the short-distance
suppression. This implies that the dibaryon pair tends to
have a small invariant mass \cite{HouSoni}.

Second, the Dalitz plot of $B\to p\bar p D^{(*)}$ \cite{ppD(star)_Babar} with asymmetric
distributions signals a nonzero angular distribution asymmetry as measured in
$B^-\to p\bar p\pi^-$, $B^-\to \Lambda_c^+\bar p\pi^+$, $B^-\to \Lambda\bar p\gamma$, $B^-\to p\bar p K^-$, and $B^-\to\Lambda\bar p\pi^-$
\cite{ppKpi_Belle,LambdaCpbarpi_Belle,Lambdapbargamma_Belle,AD_Belle,ADLambdapbarpi_Belle}.
Nonetheless, while the above-mentioned short-distance picture predicts a
correct angular distribution pattern for the decays $B^-\to p\bar
p\pi^-$, $B^-\to \Lambda_c^+\bar p\pi^+$, and  $B^-\to \Lambda\bar
p\gamma$, it fails to explain the angular correlation observed in
$B^-\to p\bar p K^-$ and $B^-\to\Lambda\bar p\pi^-$.
The intuitive argument that the $K^-$ in the $p\bar p$ rest frame is expected to emerge
parallel to $\bar p$ is not borne out by experiment
\cite{HYreview}. Likewise, the naive argument that the pion has
no preference for its correlation with the $\Lambda$ or the $\bar
p$ in the decay $B^-\to\Lambda\bar p\pi^-$ is ruled out by the new
Belle experiment \cite{ADLambdapbarpi_Belle} in which a strong correlation between the $\Lambda$ and
the pion is seen.

Therefore, the study of the charmful baryonic $B$ decays $B\to {\bf B \bar
B'} M_c$ may help improve our understanding of the underlying
mechanism for the threshold enhancement and the angular distribution
in three-body decays.
We shall therefore focus on $B\to {\bf
B \bar B'} M_c$ with $M_c=D^{(*)}$ or $J/\Psi$ to see if we can explain the branching fractions.

\section{Experimental Data}
\begin{table}[h!]
\tbl{Branching ratios of $B\to {\bf B\bar B'}M_c$ decays in units of $10^{-4}$ for $M_c=D^{(*)}$ and $10^{-6}$ for $M_c=J/\Psi$.}
{\begin{tabular}{@{}llll@{}}
\toprule
Decay & CLEO \cite{Dstarpn_Cleo}& BaBar \cite{ppD(star)_Babar,JLambdapbar_Babar}  & Belle \cite{ppD(star)_Belle,DsLambdapbar_Belle,JLambdapbar_Belle}\\
\colrule
$\bar B^0\to n\bar p D^{*+}$                &$14.5^{+3.4}_{-3.0}\pm 2.7$&                       &\\
$\bar B^0\to p\bar p D^{0}$                 &                           &$1.13\pm 0.06 \pm 0.08$&$1.18\pm 0.15 \pm 0.16$\\
$\bar B^0\to p\bar p D^{*0}$                &                           &$1.01\pm 0.10 \pm 0.09$&$1.20^{+0.33}_{-0.29}\pm 0.21$\\
$ B^-\to p\bar p D^{-}$                     &                           &                       &$<0.15$\\
$ B^-\to p\bar p D^{*-}$                    &                           &                       &$<0.15$\\
$\bar B^0\to \Lambda\bar p D^{+}_s$         &                           &                       &$0.29\pm 0.07\pm 0.05\pm 0.04$\\
$\bar B^0\to  \Lambda\bar \Lambda D^0$      &                           &                       &$(1.05^{+0.57}_{-0.44}\pm 0.14<2.6)\times 10^{-1}$\\
\hline
$B^-\to \Lambda\bar p J/\Psi$               &                           &$11.6^{+7.4+4.2}_{-5.3-1.8}$&$11.6\pm 2.8^{+1.8}_{-2.3}$\\
$B^-\to \Sigma^0\bar p J/\Psi$              &                           &&$<11$\\
$\bar B^0\to  p\bar p J/\Psi$               &                           &$<1.9$&$<0.83$\\
\botrule
\end{tabular}}
\label{table1}
\end{table}
The branching fractions of the charmful baryonic $B\to {\bf B\bar B'}M_c$ decays are summarized in Table \ref{table1},
where $\bar B^0\to n\bar p D^{*+}$ was first observed by CLEO in 2001 \cite{Dstarpn_Cleo}.
Note that the decays $\bar B^0\to p\bar p D^{0}$ and $\bar B^0\to p\bar p D^{*0}$ have similar results in rates.
The nonobservation of $B^-\to p\bar p D^{(*)-}$ is due to the fact that it proceeds via $b\to u\bar c d$ at the quark level.
This leads to a suppression of $|V_{ub} V_{cd}^*/V_{cb}V_{ud}^*|^2\simeq 10^{-4}$ compared to its neutral partner.
Likewise, it is expected that ${\cal B}(\bar B^0\to p\bar p J/\Psi)=|V_{cd}/V_{cs}|^2\,{ Br}(B^-\to \Lambda\bar p
J/\Psi)\simeq 10^{-7}$, consistent with the experimental upper bound for this decay mode.
As for the rate difference between $\bar B^0\to p\bar p
D^{0}$ and $\bar B^0\to \Lambda\bar \Lambda D^{0}$,
it has to do with the baryonic form
factors, which we are going to elaborate on later.
\section{Formalism}
To have the amplitudes, we shall adopt the generalized factorization approach \cite{Hamiltonian,ali},
which has been well applied to the study of three-body baryonic $B$ decays
\cite{HYpole,HYradi2,HYchamBaryon,HYreview,HYchamBaryon2,ChuaHouTsai0,ChuaHouTsai,
ChuaHouTsai2,ChuaHou,Tsai,Geng,AngdisppK,CP_ppKstar,ppKstar_Geng,NF_Geng,BBgamma2_ChengYang,CPandT}.
\begin{figure}[t!]
\centering
\includegraphics[width=2.8in]{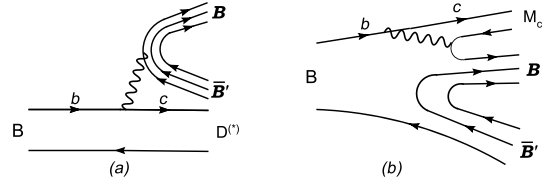}
\caption{Two types of the $B\to {\bf
B \bar B'} M_c$ decay process: (a) current type and (b) transition type.}\label{fig}
\end{figure}
Under the factorization approximation, the decay amplitudes can be classified into three different categories: the current-type (class-I),
the transition-type (class-II), and the hybrid-type (class-III) amplitudes.
The amplitude of $\bar B^0\to n\bar p D^{*+}$ is the class-I current-type which proceed via a
color-allowed, external $W$-emission diagram as depicted in Fig. \ref{fig}(a), which is given by
\begin{eqnarray}\label{AC}
{\cal A_C}(\bar B^0\to n\bar p D^{*+})&=&\frac{G_F}{\sqrt 2}V_{cb}V_{ud}^*a_1^{D^{*}}
\langle n\bar p|(\bar d u)_{V-A}|0\rangle \langle D^{*+}|(\bar c b)_{V-A}|\bar B^0\rangle\,,
\end{eqnarray}
with $a_1^{D^{*}}$ to be specified later.
The amplitudes of 
$\bar B^0\to p\bar p D^{(*)0}$, $\bar B^0\to \Lambda\bar \Lambda D^0$, 
$B^-\to \Lambda\bar p(\Sigma^0\bar p) J/\Psi$, and $\bar B^0\to p\bar p J/\Psi$ 
are the class-II transition-type via the color-suppressed internal $W$ emission diagram [Fig. \ref{fig}(b)].
The amplitudes read
\begin{eqnarray}\label{AT}
{\cal A_T}(B\to {\bf B\bar B'}M_c)&=&\frac{G_F}{\sqrt 2}V_{cb}V_{qq'}^*a_2^{M_c}\langle M_c|(\bar c q)_{V-A}|0\rangle\langle{\bf B\bar B'}|(\bar q' b)_{V-A}|B\rangle\,,
\end{eqnarray}
with $a_2^{M_c}$ to be given later, where $B\to {\bf B\bar B'}M_c$ could be
$\bar B^0\to p\bar p D^{(*)0},\;\Lambda\bar \Lambda D^{0}$ for $qq'=ud$,
$\bar B^0\to p\bar p J/\Psi$ for $qq'=cd$,
$B^-\to\Lambda\bar p J/\Psi$, $\Sigma^0\bar p J/\Psi$ for $qq'=cs$.\\
For the dibaryon creation in Eq. (\ref{AC}), we write
\begin{eqnarray}\label{timelikeF}
\langle {\bf B}{\bf\bar B'}|\bar q_1\gamma_\mu q_2|0\rangle
&=& \bar u\bigg\{[F_1(t)+F_2(t)]\gamma_\mu+\frac{F_2(t)}{m_{\bf B}+m_{\bf \bar B'}}(p_{\bf \bar B'}-p_{\bf B})_\mu\bigg\}v\;,\nonumber\\
\langle {\bf B}{\bf\bar B'}|\bar q_1\gamma_\mu \gamma_5 q_2|0\rangle
&=&\bar u\bigg\{g_A(t)\gamma_\mu+\frac{h_A(t)}{m_{\bf B}+m_{\bf \bar B'}}q_\mu\bigg\}\gamma_5 v\,,
\end{eqnarray}
where $u$($v$) is the (anti-)baryon spinor, and $F_{1,2}$,
$g_A$, $h_A$ are timelike baryonic form factors. Note that there are two additional form factors in the form of $\bar uq_\mu v$ and $\bar u\sigma_{\mu\nu}q_\nu\gamma_5 v$. However, since we assume SU(3) flavor symmetry, we can neglect these two form factors as they vanish for conserved currents. The asymptotic behavior of form factors is governed by the pQCD
counting rules \cite{Brodsky1,Brodsky3}. In the large $t$ limit,  the
momentum dependence of the form factors $F_1(t)$ and $g_A(t)$ behaves as $1/t^2$ as there are two hard gluon exchanges between the valence quarks. More precisely, in the $t\to\infty$ limit
\begin{eqnarray}\label{timelikeF2}
F_1(t)=\frac{C_{F_1}}{t^2}\bigg[\text{ln}\bigg(\frac{t}{\Lambda_0^2}\bigg)\bigg]^{-\gamma}\;, \qquad g_A(t)=\frac{C_{g_A}}{t^2}\bigg[\text{ln}\bigg(\frac{t}{\Lambda_0^2}\bigg)\bigg]^{-\gamma}\;,
\end{eqnarray}
where $\gamma=2+4/(3\beta)=2.148$ with $\beta$ being the QCD $\beta$ function and $\Lambda_0=0.3$ GeV.
In the asymptotic $t\to\infty$ limit, both $F_2(t)$ and $h_A(t)$ have an extra $1/t$ dependence relative to $F_1$ and $g_A$ owing to a mass insertion at the quark line \cite{Tsai,F2,F2b}. However,
the form factor $h_A$  is related to $g_A$ by the relation of
$h_A=-g_A{(m_{\bf B}+m_{\bf \bar B'})^2}/{t}$,
through the equation of motion. Hence, in ensuing numerical analysis we will keep $h_A(t)$ and neglect $F_2(t)$.
Under the SU(3) flavor and SU(2) spin symmetries \cite{Brodsky3}, the parameters $C_{F_1}$ and
$C_{g_A}$  appearing in $0\to {\bf B\bar B'}$ transitions are no longer
independent but are related to each other through the two reduced parameters $C_{||}$
and $C_{\overline{||}}$. Then we have
\begin{eqnarray}
C_{F_1}=\frac{4}{3}C_{||}-\frac{1}{3}C_{\overline{||}}\,\;\;,C_{g_A}=\frac{4}{3}C_{||}+\frac{1}{3}C_{\overline{||}}\,,\;\;\text{for $0\to n\bar p$},
\end{eqnarray}
As for the three-body transition $B\to {\bf B\bar B'}$, its most general expression reads
\begin{eqnarray}\label{transitionF}
&&\langle {\bf B}{\bf\bar B'}|V_\mu|B\rangle=i\bar u[  g_1\gamma_{\mu}+g_2i\sigma_{\mu\nu}p^\nu +g_3p_{\mu} +g_4(p_{\bf\bar B'}+p_{\bf B})_\mu +g_5(p_{\bf\bar B'}-p_{\bf B})_\mu]\gamma_5v,\nonumber\\
&&\langle {\bf B}{\bf\bar B'}|A_\mu|B\rangle=i\bar u[ f_1\gamma_{\mu}+f_2i\sigma_{\mu\nu}p^\nu +f_3p_{\mu} +f_4(p_{\bf\bar B'}+p_{\bf B})_\mu +f_5(p_{\bf\bar B'}-p_{\bf B})_\mu]        v,
\end{eqnarray}
with $V(A)_\mu=\bar q'\gamma_\mu (\gamma_5) b$ and $p=p_B-p_{\bf B}-p_{\bf\bar B'}$.
Since three gluons are needed to induce the $B\to{\bf B\bar B'}$ transition,
two for producing the baryon pair and one for kicking the spectator quark in the $B$ meson,
the pQCD counting rules imply that to the leading order
\begin{eqnarray}\label{transitionF2}
f_i(t)=\frac{D_{f_i}}{t^3}\;, \qquad g_i(t)=\frac{D_{g_i}}{t^3}\;.
\end{eqnarray}
Just as the previous case for vacuum  to the dibaryon transition, under the SU(3) flavor and $SU(2)$ spin symmetries, the parameters $D_{g_1}$, $D_{f_1}$, $D_{g_i}$ and $D_{f_i}$ can be expressed in terms of the reduced parameters $D_{||}$, $D_{\overline{||}}$ and $D^i_{||}$.
Then we have
\begin{eqnarray}\label{D0}
&D_{g_1}^{p\bar p}(D_{f_1}^{p\bar p})=\frac{1}{3}D_{||}\mp\frac{2}{3}D_{\overline{||}},\;\;D_{g_i}^{p\bar p}=-D_{f_i}^{p\bar p}=-\frac{1}{3}D^i_{||},&\;\;\text{for $\bar B^0\,\to p\bar p$},\nonumber\\
&D_{g_1}^{\Lambda\bar \Lambda}(D_{f_1}^{\Lambda\bar \Lambda})=\frac{1}{2}D_{||}\mp\frac{1}{2}D_{\overline{||}},\;\;D_{g_i}^{\Lambda\bar \Lambda}=D_{f_i}^{\Lambda\bar \Lambda}=0,\;\;\;\;\;\;\;\;\;\;\;&\;\;\text{for $\bar B^0\,\to \Lambda\bar \Lambda$},\nonumber\\
&D_{g_1}^{\Lambda\bar p}(D_{f_1}^{\Lambda\bar p})=-\sqrt{\frac{3}{2}}D_{||},\;\;\;D_{g_i}^{\Lambda\bar p}=-D_{f_i}^{\Lambda\bar p}=-\sqrt{\frac{3}{2}}D^i_{||},&\;\;\text{for $\bar B^-\to \Lambda\bar p$},
\end{eqnarray}
where the upper (lower) sign is for $D_{g_1}$ ($D_{f_1}$) with $i=2, 3, ..., 5$.
For the meson parts, the values for the decay constants of $D^{(*)0}$ and $J/\Psi$ can be found in Refs. \refcite{pdg,fDstar}.
The definition and parametrization of $B$ to $D^{(*)}$ transition form factors can be found in Refs. \refcite{BSW,BtoD}.

\section{Numerical Analysis}
We need to specify various input parameters for a numerical analysis.
For the CKM matrix elements, we use the values of the Wolfenstein parameters in Ref. \refcite{CKMfitter}.
Here, we extract $a_1^{D^{*}}$, $a_2^{D^{(*)}}$, and $a_2^{J/\Psi}$  from
$\bar B^0\to n\bar p D^{*+}$, $\bar B^0\to p\bar p D^{*0}$, and $B^-\to \Lambda\bar p J/\Psi$, respectively,
which are given by
\begin{eqnarray}\label{a1a2}
a_1^{D^*}=1.23\pm 0.19\;,\;\;a_2^{D^{(*)}}=0.33\pm 0.04\;,\;\;a_2^{J/\Psi}=0.17\pm 0.03\;.
\end{eqnarray}
We note that factorization works if the parameters $a_1$ and $a_2$ are universal;
namely, they are channel by channel independent.
Since $a_i^{M_c}$ lie in the ranges of $a_1\sim {\cal O}(1)$ and $a_2\sim {\cal O}(0.2-0.3)$ \cite{ai},
which is suggested by two-body mesonic $B$ decays, we shall assume the validity of factorization in charmful baryonic $B$ decays.

\begin{figure}[h!]
\centering
\includegraphics[width=1.4in]{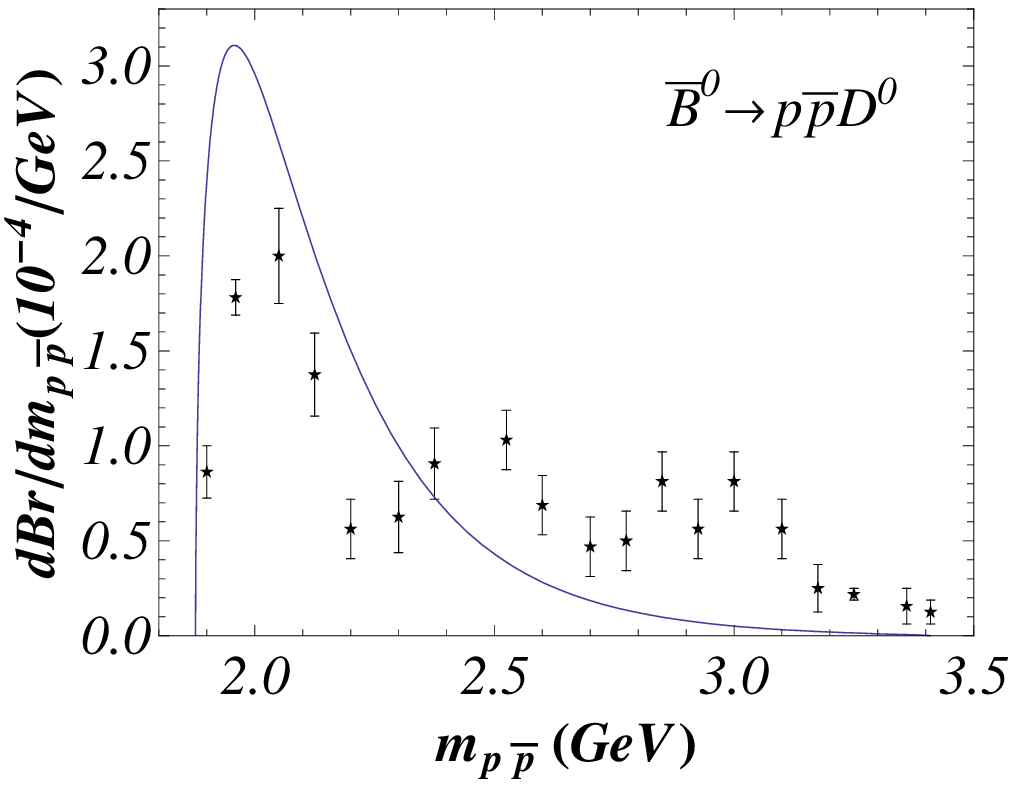}
\includegraphics[width=1.4in]{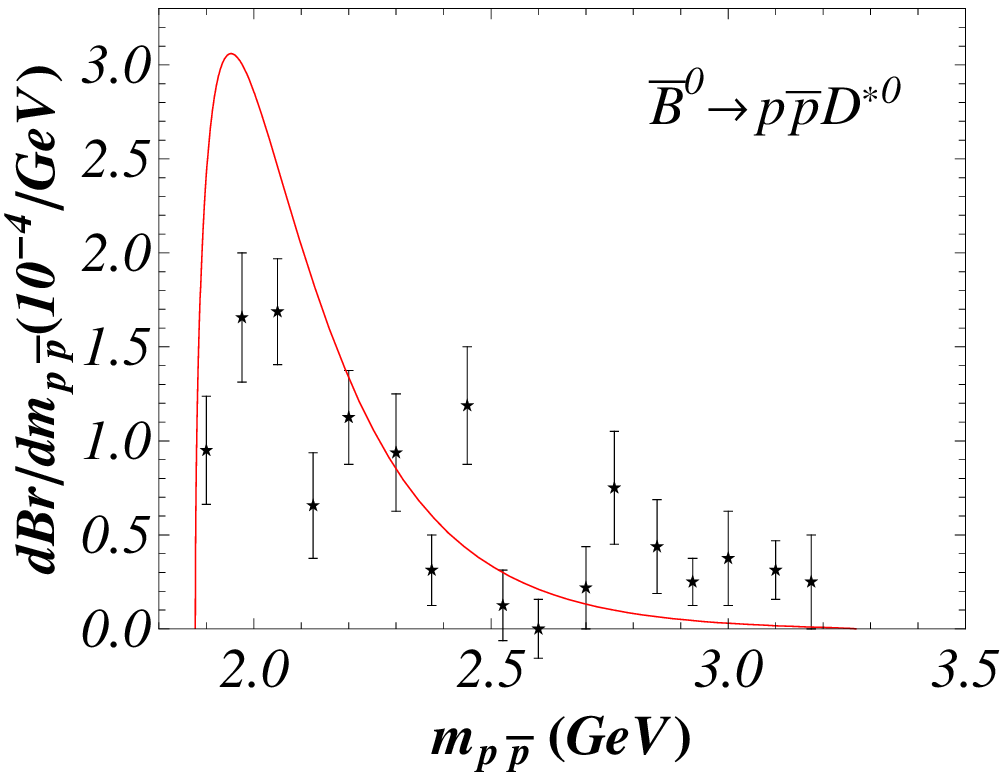}
\includegraphics[width=1.45in]{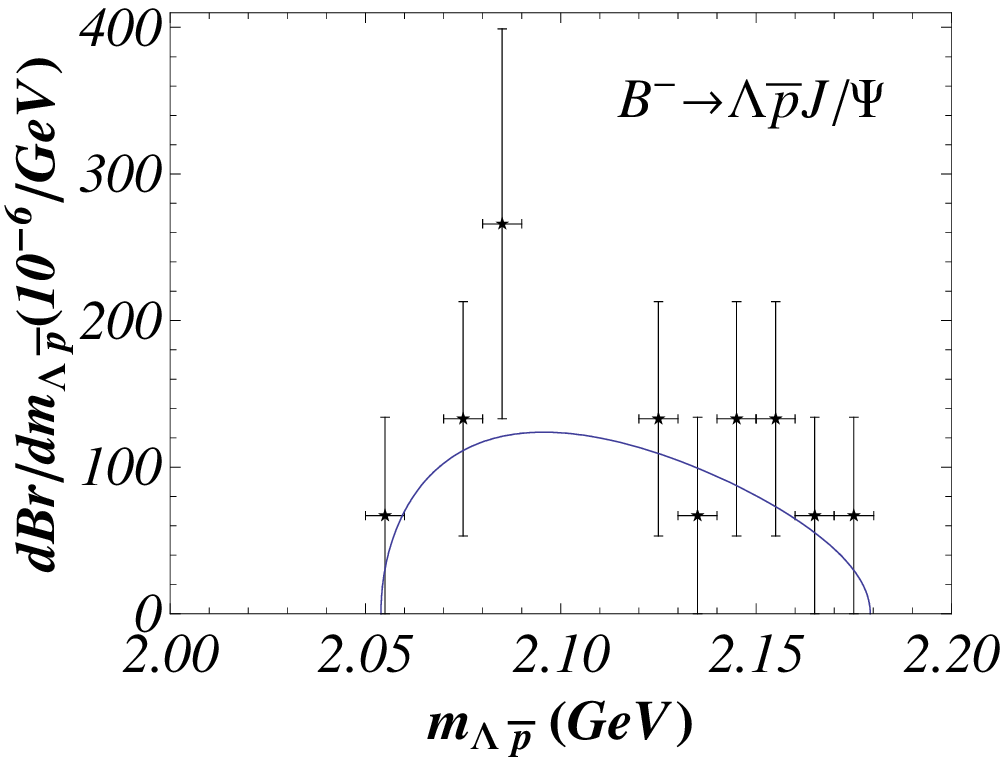}
\caption{Dibaryon invariant mass distributions for $\bar B^0\to
p\bar p D^{0}$, $\bar B^0\to
p\bar p D^{*0}$ and $B^-\to \Lambda \bar p J/\Psi$, respectively.
Experimental data are taken from Refs. \refcite{ppD(star)_Babar,JLambdapbar_Belle}.}\label{fig_threshold}
\end{figure}

For the parameters $C_{||}$ and $C_{\overline{||}}$  in
Eqs. ({\ref{timelikeF}, \ref{timelikeF2}}), we use the data of  $e^+ e^-\to p\bar
p,\;n\bar n$ \cite{eetopp,eetonn} to determine their magnitudes
and the decay rate of $\bar B^0 \to n\bar p D^{*+}$ to fix their relative sign
\begin{eqnarray}\label{C1}
&&(C_{||},\,C_{\overline{||}})=(67.9\pm 1.4,\,-216.9\pm 23.5)\,{\rm GeV^{4}}\;.
\end{eqnarray}
As for the parameters $D_{||}$ and $D_{\overline{||}}$ in Eqs. (\ref{transitionF},
\ref{transitionF2}),  we employ the observed rates of $\bar B^0 \to p\bar p D^{0}$,
$B^- \to p\bar p K^{*-}$, $\bar B^0 \to p\bar p K^{*0}$, and $B^-
\to p\bar p \pi^-$
\cite{ppK(star)pi_Belle,AD_Belle,ppKpi_Belle,pppi_Babar,ppD(star)_Belle,ppD(star)_Babar,ppKstar_Belle}
in conjunction  with the measured angular distribution of the last decay mode to obtain
\begin{eqnarray}\label{D1}
&&(D_{||},\;D_{\overline{||}})=(67.7\pm 16.3,\,-280.0\pm 35.9)\;{\rm GeV^5},\nonumber\\
&&(D_{||}^2,\,D_{||}^3,\,D_{||}^4,\,D_{||}^5)=\nonumber\\
&&(-187.3\pm 26.6,\,-840.1\pm 132.1,\,-10.1\pm 10.8,\,-157.0\pm 27.1)\;{\rm GeV^4}\;.
\end{eqnarray}
To calculate the decay rates, we use the equation in Ref. \refcite{pdg} for three-body decays.
We present the numerical results for the branching ratios in Table \ref{table2}, and the dibaryon invariant mass spectrum for
$\bar B^0\to p\bar p D^{(*)0}$ and $B^-\to \Lambda\bar p J/\Psi$ in Fig. \ref{fig_threshold}.
We note that the first and second errors in Table \ref{table2} come from the uncertainties of $a_1$ ($a_2$) in Eq. (\ref{a1a2})
and baryonic form factors in Eqs. (\ref{C1} ,\ref{D1}), respectively.
\begin{table}[t!]
\tbl{Branching ratios of $B\to {\bf B\bar B'}M_c$ decays,
where the first and second errors come from the uncertainties of $a_1$ ($a_2$) in Eq. (\ref{a1a2})
and baryonic form factors in Eqs. (\ref{C1} ,\ref{D1}), respectively.}
{\begin{tabular}{@{}lclc@{}}
\toprule
${\cal B}\times 10^{-4}$&Our work &${\cal B}\times 10^{-6}$&Our work\\
\colrule
$\bar B^0\to n\bar p D^{*+}$                &$14.4\pm 4.8\pm 3.2$  &$\bar B^0\to  \Lambda\bar \Lambda D^0$  &$2.3\pm 0.6\pm 0.6$\\
$\bar B^0\to p\bar p D^{0}$                 &$1.1 \pm 0.3\pm 0.2$  &$B^-\to \Lambda\bar p J/\Psi$           &$11.6\pm 4.5\pm 2.9$\\
$\bar B^0\to p\bar p D^{*0}$                &$1.0\pm0.3\pm 0.4$    &$B^-\to \Sigma^0\bar p J/\Psi$          &$0.13\pm 0.05\pm 0.02$\\
                                            &                      &$\bar B^0\to  p\bar p J/\Psi$           &$1.2\pm 0.5\pm 0.2$\\
                                            \botrule
\end{tabular}}
\label{table2}
\end{table}

\section{Discussion and Conclusion}
As seen in Table \ref{table2}, our prediction ${\cal B}(B^-\to \Sigma^0\bar p J/\Psi)=1.3\times 10^{-7}$ is consistent with the Belle limit, $1.1\times 10^{-5}$, and ${\cal B}(\bar B^0\to p\bar p J/\Psi)=1.2\times 10^{-6}$ is in accordance with the BaBar limit but slightly higher than the upper bound set by Belle. As for the threshold peaking effect,
while it manifests in the decay $\bar B^0\to p\bar p D^{(*)0}$ the data clearly do not show the threshold behavior in $B^-\to \Lambda\bar p J/\Psi$ (see Fig. \ref{fig_threshold}).
This can be understood as follows.
In the latter decay, the invariant mass $m_{\Lambda\bar p}$ ranges from 2.05 to 2.18 GeV, which is very narrow compared to the $m_{p\bar p}$ range in the $\bar B^0\to p\bar p D^{(*)0}$ decay. Consequently, the invariant mass distribution of $d\Gamma/dm_{\Lambda\bar p}$
is governed by the shape of the phase space due to the relative flat $1/t^3$ dependence within the small allowed $m_{\Lambda\bar p}$ region.
${\cal B}(\bar B^0\to \Lambda\bar \Lambda D^0)$ predicted to be of order $2.3\times 10^{-6}$ is consistent with the measured Belle data, $(1.05^{+0.57}_{-0.44}\pm 0.14)\times 10^{-5}<2.6\times 10^{-5}$.
${\cal B}(\bar B^0\to \Lambda\bar \Lambda D^0)<{\cal B}(\bar B^0\to p\bar p D^0)$ can be understood by the baryonic form factors in Eqs. (\ref{D0}, \ref{D1}), which are evaluated to be
\begin{eqnarray}
&&(D_{g_1}^{\Lambda\bar \Lambda},\;D_{f_1}^{\Lambda\bar \Lambda})=(172.8,\;-105.1)\;\text{GeV}^5
\,,\;(D_{g_1}^{p\bar p},\;D_{f_1}^{p\bar p})=(207.9,\;-162.7)\;\text{GeV}^5\,,\nonumber\\
&&(D^{\Lambda\bar \Lambda}_{g_i},D^{\Lambda\bar \Lambda}_{f_i})=0\,,\;(D^{p\bar p}_{g_i},\;D^{p\bar p}_{f_i})\neq 0\,.\;(\text{i=2, 3, ..., 5})
\end{eqnarray}

In sum, within the framework of  the generalized factorization approach,
we have explained the measured charmful three-body baryonic $B$ decays with
$D^{(*)}$ or $J/\Psi$ in the final state.
The measured ${\cal B}(\bar B^0\to \Lambda\bar \Lambda D^0)<{\cal B}(\bar B^0\to p\bar p D^0)$
can be understood. This is due to the constraint in the baryonic transition form factors for $\bar B^0\to \Lambda\bar \Lambda$
in the approach of the pQCD counting rules.

\section*{Acknowledgments}
The author is very grateful to Hai-Yang Cheng, Chao-Qiang Geng, and Chun-Hung Chen for the fruitful collaboration
on the charmful three-body baryonic B decays and to Min-Zu Wang for useful discussions.

\end{document}